\documentclass[fleqn,usenatbib]{mnras}

\usepackage{newtxtext,newtxmath}
\usepackage[T1]{fontenc}

\DeclareRobustCommand{\VAN}[3]{#2}
\let\VANthebibliography\thebibliography
\def\thebibliography{\DeclareRobustCommand{\VAN}[3]{##3}\VANthebibliography}

\usepackage{graphicx}	
\usepackage{amsmath}	
\usepackage{mathtools}
\usepackage{mathrsfs}
\usepackage{enumerate}
\usepackage{morefloats}
\usepackage{longtable}
\usepackage{afterpage}
\usepackage{acronym}   
\hypersetup{linkcolor=red,citecolor=blue,urlcolor=magenta}
\usepackage{enumitem}
\usepackage{soul}

\usepackage[makeroom]{cancel}
\newcommand*{\XPSI}{X-PSI\xspace}
\newcommand*{\NICER}{{NICER}\xspace}

\newcommand*{\MultiNest}{\textsc{MultiNest}\xspace}

\acrodef{MSP}{millisecond pulsar}
\acrodef{PPM}{Pulse Profile Modelling}
\acrodef{EoS}{equation of state}
\acrodef{ISS}{International Space Station}
\acrodef{NS}{neutron star}
\acrodefplural{NS}[NSs]{neutron stars}
\graphicspath{{figs/}}

\usepackage[dvipsnames]{xcolor}
\usepackage{xspace}

\title[Spin rate and pulse profile modelling]{Scaling relations for the uncertainty in neutron star radius
inferred from pulse profile modelling: the effect of spin rate}

\author[E. Bootsma et al.]{
Erik Bootsma,$^{1}$
Serena Vinciguerra,$^{1}$
Anna L. Watts\thanks{E-mail: A.L.Watts@uva.nl},$^{1}$
Yves Kini$^{1}$
and Tuomo Salmi$^{1,2}$
\\
$^{1}$Anton Pannekoek Institute for Astronomy, University of Amsterdam, P.O. Box 94249, 1090GE Amsterdam, the Netherlands\\
$^{2}$Department of Physics, P.O. Box 64, FI-00014 University of Helsinki, Finland\\
}

\date{Accepted XXX. Received YYY; in original form ZZZ}

\pubyear{2025}

\begin{document}
\label{firstpage}
\pagerange{\pageref{firstpage}--\pageref{lastpage}}
\maketitle

\begin{abstract}
Pulse profile modelling using X-ray data from \NICER permits the inference of mass and radius for rotation-powered millisecond pulsars. This in turn constrains the equation of state of cold dense matter.  Previous studies indicate that the uncertainty in the inferred radius should reduce as neutron star spin rate increases. 
Here we test this using one of the pipelines currently being used for pulse profile modelling with \NICER data. 
We synthesize a set of pulse profiles, assuming different neutron star spin frequencies, spanning the range (25-700)\,Hz. All of the simulated data sets are generated with the same (single) hot spot configuration, assuming a neutron star mass and radius of $1.6\,M_{\mathrm{\odot}}$ and $10\,$km. 
For this restricted set of synthetic data, we find no improvement in the radius credible interval once spin frequency exceeds a certain value (in this specific  case $\sim 200$\,Hz).  If this result were to apply more generally, it would have important implications for the observing strategy for current and future pulse profile modelling missions: 
targets can be prioritized based on properties other than their spin frequencies, as long as we are in the millisecond range. 
\end{abstract}


\section{Introduction}
In the last few years, X-ray data from \NICER \citep[the Neutron Star Composition ExploreR,][]{Gendreau2016} 
has enabled inference of the mass and radius of a set of neutron stars \citep{Riley2019,Miller2019,Riley2021,Miller2021,Salmi2022,Salmi2023,Vinciguerra2023b,Salmi24,Dittmann24,Choudhury24,Salmi24b} 
providing new constraints on the \ac{EoS} of cold dense matter \citep[see e.g.,][]{Annala2023,Takatsy23,Zhu23,Huang24,Pang24,Kurkela24,Rutherford24}. 

This has been achieved through X-ray observations of rotation-powered \acp{MSP}. 
The X-ray emission of \acp{MSP} derives from return currents that heat up
the pulsar's surface at the magnetic poles.
The data collected by \NICER have been analysed with \ac{PPM}, a relativistic ray-tracing technique that enables inference of the pulsar's mass ($M$) and equatorial radius ($R_{\mathrm{eq}}$) by modelling the periodic modulation of its X-ray flux as a function energy and time, i.e. its pulse profile \citep[see e.g.][]{Poutanen03,Poutanen2006,Cadeau07, Morsink2007,Psaltis2014,AlGendy2014,Bogdanov2019b}.
Over the next few years \NICER is expected to deliver more and improved measurements at the $\sim \pm 10\%$ level (68\% credible interval).  
However future missions such as NewAthena \citep{NewAthena}, STROBE-X \citep{strobex} or eXTP \citep{Watts2019eXTP} could provide much tighter constraints.
Time on these facilities will be in high demand, and \ac{PPM} is a computationally expensive task. It will be critical 
to focus the available resources on the most promising sources. \\

Theoretically we expect tighter radius constraints for the fastest spinning \acp{NS}, as special relativistic effects, rising with high rotational speeds, can help disentangle the effects of equatorial radius from those of the compactness $C$ ($C = GMR_{\mathrm{eq}}^{-1}c^{-2}$ with $c$ the speed of light and $G$ the universal gravitational constant), to which \ac{PPM} is particularly sensitive \citep[see][and references therein]{Bogdanov2019b}. 
Earlier studies, based on analytic approximations \citep{Poutanen2006,Psaltis2014} or simulations \citep{Lo2013, MillerLamb2015,Bogdanov2021}, appear to confirm this prediction, implying that resources might sensibly be prioritized towards more rapidly-spinning sources. 

In this paper we test the scaling of radius uncertainties with neutron star spin rate using \XPSI \citep[X-ray Pulse Simulation and Inference, ][]{Riley2023}, one of the full \ac{PPM} inference pipelines currently being used in the analysis of \NICER \ac{MSP} data. We also compare our findings with the theoretical expectations for the spin scaling, building on work by \citet{Psaltis2014}.  The simulation and inference process is described fully in Section \ref{sec:methodology}.
We present our findings in Section \ref{sec:results} and discuss them in Section \ref{sec:discussion}. We summarise our study in Section \ref{sec:conclusions}.

\section{Methodology}
\label{sec:methodology}

In Section \ref{subsec:prediction} we summarise some predictions for the scaling of uncertainties with spin rate from \citet{Psaltis2014}. We then set up to test this using one of the current \ac{PPM} inference pipelines, \XPSI \citep[][see also the GitHub repository \url{https://github.com/xpsi-group/xpsi}]{Riley2023}.  
To both generate our synthetic data and to analyse them, we used \XPSI version \texttt{0.7.8}. 
In Section \ref{subsec:xpsi}, we briefly outline the main steps performed in \XPSI that are relevant to this work \citep[for more detailed explanations see][]{riley_thesis,Riley2019,Riley2021, Vinciguerra2023a}.

\NICER collects counts per pulse-invariant (PI) channel (nominally associated with specific energy intervals) and associates a time with each of them. Thanks to radio measurements, the spin frequency of all \NICER targets for \ac{PPM} is well known. 
The data registered by \NICER over many rotation cycles is then folded (since for \NICER targets we do not expect the hot spot properties to change), according to the specific rotation period  of the \ac{MSP} of interest.  The pulse profiles, analysed with \XPSI, therefore have the final format of counts per PI channel per phase bin, where the cycle is typically divided into 32 bins.  For this study we focus on PI channels in the range $[30,300)$, corresponding to nominal energies $[0.3,3.0)\,$keV \footnote{The bulk of the simulated emission falls indeed well within this range, as shown in Figure \ref{fig:simulated}. This cut, compared to the whole energy range of sensitivity of \NICER allows a considerable reduction of the computational resources necessary for inference analyses. }. We present our synthetic pulse profile data sets in Section \ref{subsec:simulations} and the inference analyses performed on them in Section \ref{subsec:inferences}.

\subsection{Prediction for scaling relations}
\label{subsec:prediction}
\citet{Psaltis2014} investigate the properties of observables required to infer mass and radius from a pulse profile produced by a \ac{NS} rotating at a known period. 
In particular, they state that for small hot spot sizes \citep{Bogdanov2007} the pulse profile generated by a single circular emitting hot spot can be determined by only mass $M$, equatorial radius $R_{\mathrm{eq}}$, observer inclination with respect to the rotation axis $i$ and colatitute of the hot spot $\theta_s$. 
However only the relativistic effects introduced by a moderate spin frequency $f$ can disentangle different mass and radius values
from the compactness. 
This is because higher spin frequencies yield larger amplitudes of the second harmonic. 
The ratio between the amplitude of the fundamental $C_1$ and second harmonic $C_2$ of the bolometric pulse profile can be approximated as \citep{Poutanen2006,Psaltis2014}:
\begin{equation}
\label{eq:amplitude_ratio}
   \frac{C_2}{C_1} \approx k\left(\frac{2\pi f R_{\mathrm{eq}}}{c}\right) \sin i\sin\theta_s,
\end{equation} 
where $k$ is a constant that contains information about the emission spectrum \citep{Poutanen2006}. As in \citet{Psaltis2014}, hereafter we consider $k=2$. 
\citet{Psaltis2014} then use this relation to infer the relative uncertainty of the inferred radius as a function of the other parameters.  Considering, for example, only the uncertainty in the second harmonic $\Delta C_2$, they show that:
\begin{equation}
\label{eq:relR}
    \frac{\Delta R_{\mathrm{eq}}}{R_{\mathrm{eq}}} = \left[ 2 C_1 \left( \frac{2 \pi f R_{\mathrm{eq}}}{c} \right) \sin{i} \sin{\theta_s} \right]^{-1} \Delta C_2.
\end{equation}
They also demonstrate that the uncertainty over any harmonic amplitude $\Delta C_n$ can be estimated as a 
function of source counts $S$ and background counts $B$:
\begin{equation}
\label{eq:DeltaCn}
\Delta C_n = \frac{\sqrt{S+B(1+C_n^2)}}{S},    
\end{equation}
where the term in parentheses under the square root appears when accounting for a Poisson noise uncertainty on the background measurement ($\Delta B = \sqrt{B}$). At first glance, Equation \ref{eq:relR}~would suggest: 
\begin{equation}
\label{eq:PsaltisFit}
\Delta R_{\mathrm{eq}}\left(f\right) \approx \alpha f^{-1}, 
\end{equation} 
where $\alpha$ is the constant of proportionality, in principle as derived in Equation \ref{eq:relR}. 
However if, from Equation \ref{eq:amplitude_ratio}, we account for the uncertainty on all four parameters (neglecting for now any potential correlations), we would get:
\begin{equation}
\label{eq:DRR}
     \Delta R_{\mathrm{eq}}\approx R_{\mathrm{eq}}\sqrt{\left(\frac{\Delta C_1}{C_1}\right)^2+\left(\frac{\Delta C_2}{C_2}\right)^2+\left(\cot i \Delta i \right)^2 + \left(\cot \theta_s \Delta \theta_s \right)^2}.
\end{equation}
Now, thanks to Equation \ref{eq:amplitude_ratio}, we can write $C_2$ as a function of $C_1$ and the spin frequency $f$ of the \ac{NS}. This shows that, more generally: 
\begin{equation}
\label{eq:deltaR_alpha_beta}
    \Delta R_{\mathrm{eq}}\left(f\right) \approx \sqrt{\beta +\frac{\gamma}{f^2}},
\end{equation}
where $\beta$ and $\gamma$ are proportionality constants in principle derivable from Equation \ref{eq:DRR}. 

In this work we test whether any of the approximate relations reported above can well represent the frequency dependence of the relative radius uncertainty, 
when compared to realistic mock data and analyses. In particular, we fit the
$\Delta R_{\mathrm{eq}}(f)$,
 obtained from our inference runs, with Equations \ref{eq:PsaltisFit} and \ref{eq:deltaR_alpha_beta}.
In all formulas we use the values of $f$
adopted to create the analysed data sets\footnote{Note that the spin frequency for the \NICER \acp{MSP} is well known from radio observations and therefore is considered to have no uncertainty for \ac{PPM} analyses.}. We define 
the radius uncertainty $\Delta R_{\mathrm{eq}}$
as the 68\% credible interval obtained from the corresponding inferred posterior distributions\footnote{This implies that we are approximating the posteriors as Gaussian distributions, as in Equations \ref{eq:relR} - \ref{eq:deltaR_alpha_beta} the uncertainties are derived as standard deviations.}. 

\subsection{X-PSI in a nutshell}
\label{subsec:xpsi}

\XPSI synthesises X-ray pulsations in the same format as the \NICER data described above, i.e. predicting X-ray photons per PI channel per phase bin. 
The software simulates the photons leaving the \ac{NS} surface from the assumed hot spot(s), accounting either for a black-body photosphere or a customised atmosphere \citep[a choice that can significantly impact the output, see e.g. ][]{Salmi2023}.  
The latter relies on interpolations of pre-computed tables that, depending on the assumed composition and ionisation state, describe the intensity of the radiation field as a function of photon energy, surface gravity, effective temperature and cosine of the photon emission angle \citep[for more details see Section 2.4.1 of][]{Riley2019}. 
In this work we assume a black-body photosphere for both creating and analysing the data sets. 

\XPSI also accounts for special and general relativistic effects influencing the energy, the time of arrival and the trajectories of the emitted photons. 
In particular our pulse profile simulations rely on the Oblate Schwarzschild plus Doppler approximation for the \ac{NS} space-time and oblate surface \citep[][]{Poutanen03,Poutanen2006,Cadeau07,Morsink2007,AlGendy2014}. 
The X-ray photons are then assumed to propagate from the source to \NICER.
The effect of the interstellar medium is calculated via attenuation factors, which affect the X-ray flux differently at different energies. 
This is done using the hydrogen column density value, following the \texttt{tbnew} model\footnote{\url{https://pulsar.sternwarte.uni-erlangen.de/wilms/research/tbabs/}}, used in \citet{Riley2019}. 
Finally, \XPSI simulates how \NICER would register these photons, given its corresponding instrument response \footnote{The \NICER instrument response describes how likely it is that a photon with a given energy will be detected by each of the considered PI channels.}, and in doing so it generates the final synthetic pulse profiles.

In \XPSI it is possible to define hot spot models of various complexities, describing the properties of the emitting \ac{NS} surface. Generally \NICER \ac{PPM}  analyses assume that the \ac{NS} pulsations are generated from localised hot spots characterised by a homogeneous temperature; \XPSI however also allows the possibility of adding emission from the surface surrounding these hot spots. 
In \XPSI, hot spots are described with one or two circular components. When used, the additional (overlapping) component could also emit at a uniform temperature or mask the emission of the first one, allowing shapes such as rings and crescents 
\citep[see model description in][for more details]{Vinciguerra2023a}. In this paper we focus on a single hot spot (with no additional emission from the rest of the star), described by one circular component of uniform temperature. 

The pulse profiles generated in this way can be used as synthetic data sets, for testing purposes (typically once background and Poisson noise are added) or, more often, in the process of mass and radius inferences.  
Within \XPSI, parameter estimation is then performed in a Bayesian inference framework. 
The prior distributions corresponding to each of the model parameters define the parameter space which, in \XPSI, is most commonly sampled by \MultiNest \citep{Feroz2008,Feroz2009,Feroz2019}, interfaced through PyMultiNest \citep{Buchner2014}.

In the inference process, each sample is then weighted by the posterior probability that the observed \NICER data is a Poisson realisation of the resulting synthetic pulse profile. 
For these posterior estimates, we use
a likelihood function which is marginalised over the background \citep[see ][ for more details]{Salmi2022}. 
In \XPSI the background, when not explicitly modeled as an additional component (e.g. as a power law), is captured by a series of independent variables, one per energy channel. Each of these variables is treated as a constant count rate that does not vary with the \ac{NS} rotation cycle and so accounts, in the corresponding energy interval, for all unpulsed components that do not directly derive from the emission of the \ac{NS} surface at that channel.

\subsection{Simulated Data}
\label{subsec:simulations}
To study the spin frequency dependence of the relative radius uncertainties,
we consider a series of data sets, created with the same physics and hot spot model parameters, but with varying rotation periods. 
In this way, systematic changes 
in the width of the radius posterior distributions can only be attributed to changes in the spin frequency. 

\begin{figure}
	\includegraphics[width=\columnwidth]{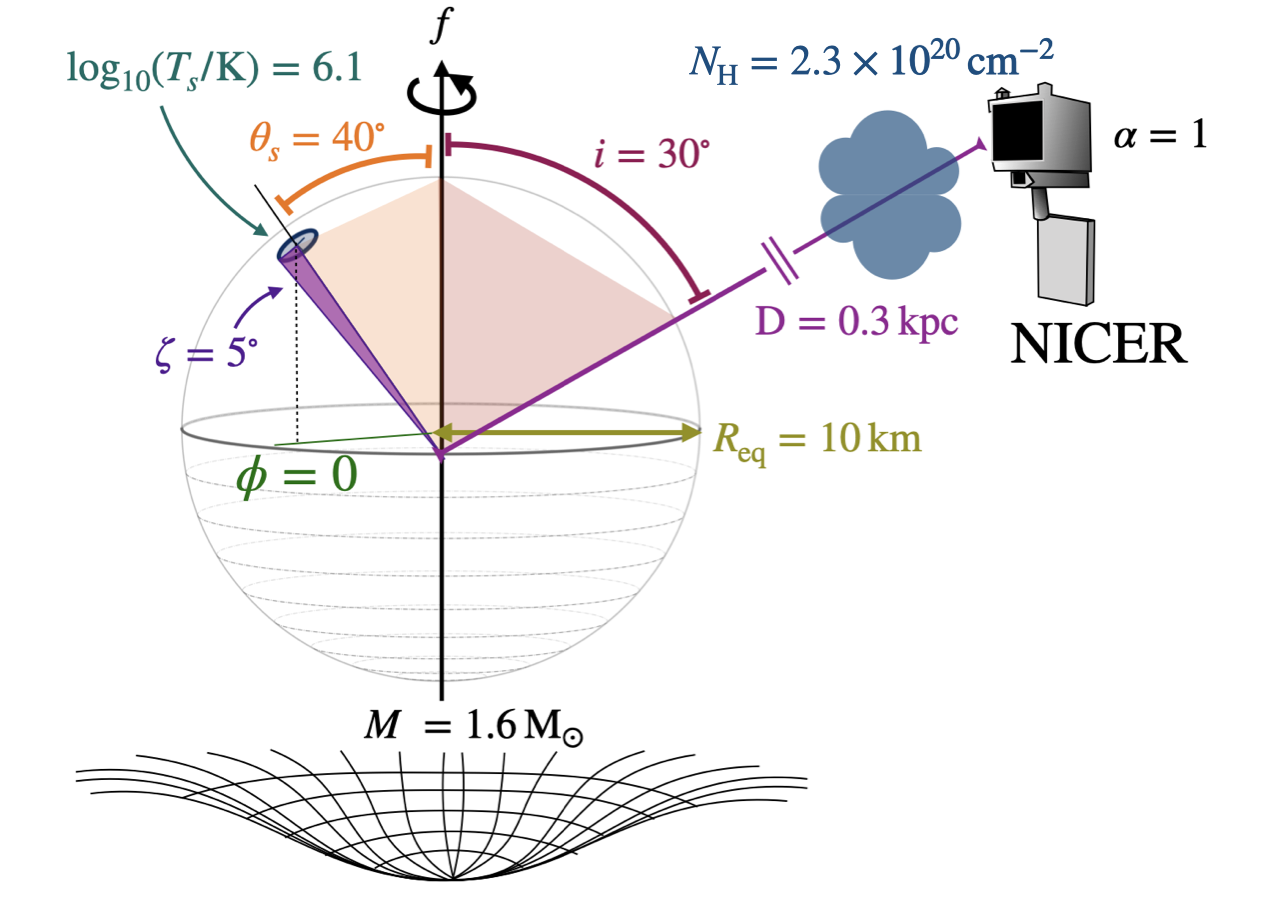}
    \caption{Schematic representation of the simulated X-ray pulsars used to generate synthetic pulse profiles. This shows both the system configuration with respect to the observer as well as the emitting surface pattern. Symbols are explained in Table \ref{tab:Parameters}. }
    \label{fig:parameters}
\end{figure}

To be as comparable as possible to the case presented in \citet{Psaltis2014}, we use \XPSI to simulate pulse profiles generated by a single hot spot, characterised by the parameters presented in Table \ref{tab:Parameters} and represented schematically in Figure \ref{fig:parameters}. 
We also need to specify some parameters that were not defined in \citet{Psaltis2014}; for distance, hot spot temperature and column density we pick values that are representative for the \NICER \acp{MSP}.
We assume that the photons produced by our simulated pulsars are registered by the \NICER XTI instrument using a representative response matrix\footnote{The response matrix used is from \citet{Salmi2022}, and is also available in the Zenodo repository for this paper \citep{zenodoScaling}.}.
We then allow for deviations by defining 
the \NICER XTI energy-independent scaling factor $\alpha_\mathrm{XTI}$, that we set to one to generate our synthetic data sets. This choice implies that the strength of the recorded emission is set by the \NICER response matrices and is not further attenuated by our choice of  $\alpha_\mathrm{XTI}$\footnote{The pulse profile is in fact only sensitive to the distance $D$ and \NICER XTI scaling factor in the combination $\alpha_\mathrm{XTI} D^{-2}$.}. 
Note that in our simulations we set source parameters, from which the observational quantities (such as $C_1$) are then derived. 
As in \citet{Psaltis2014}, we assume that the hot spot emits black-body radiation and do not account for atmospheric effects. 

We then vary the spin period of the simulated \ac{NS} to form data sets spanning the frequency range of interest: $\sim\left(\mathrm{few} - 700\right)$\,Hz. 
We sampled the frequency range $100\,\mathrm{Hz} \leq f \leq 700 \, \mathrm{Hz}$ every $100$\,Hz, to fully explore the range covered by known \acp{MSP}, and on the low end two additional frequencies $f = 50 \ \mathrm{Hz}$ and $f = 25 \ \mathrm{Hz}$.
We stop at 700\,Hz to limit the possible impact of systematic errors deriving from our use of the Oblate Schwarzschild plus Doppler approximation \citep{Morsink2007,AlGendy2014,Psaltis14_HT,Silva2021}. 

\begin{figure}
	\includegraphics[width=\columnwidth]{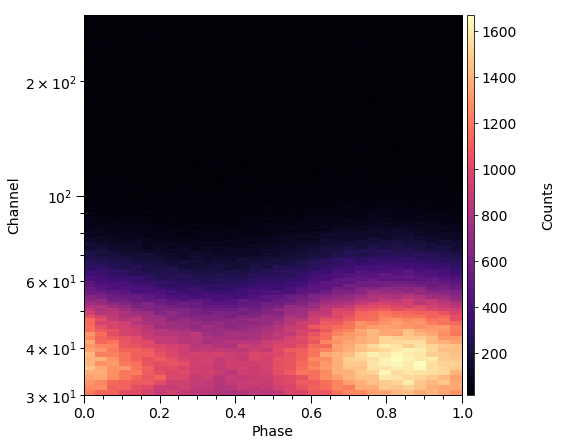}
    \caption{Synthetic Pulse Profile at spin frequency $f=600$\,Hz. The model parameter values used to create this data set are reported in Table \ref{tab:Parameters}, while the simulation procedure is outlined in Section \ref{subsec:simulations}. 
    On the y-axis is the PI channel for \NICER, where higher numbers correspond to higher nominal energies. On the x-axis are phase bins, spanning one rotation cycle. The colour indicates the X-ray counts per phase bin and channel. }
    \label{fig:simulated}
\end{figure}

The synthetic pulse profiles generated for this investigation have $10^6$ source (X-ray) counts, as considered by \citet{Psaltis2014}. 
To this we add a constant background whose total counts are set to be a third of the total source counts $B =S/3$. The background is added after we simulate how \NICER would receive the emission from the \ac{NS} and it is uniformly distributed throughout all phases and channels. 
Finally we add Poisson noise to each synthetic signal (hereafter we will use the word signal to indicate the counts resulting from summing the source emission and the background). 
Since different noise realisations, as well as the random processes occurring in the analyses (if the settings are not adequate), could yield changes in our inferred posteriors
\citep{Vinciguerra2023a}, 
we decided to simulate three different data sets for each spin frequency. 
In this way we should be able to disentangle systematic changes, coming from different rotation periods, from the differences created by the particular realisations of these random processes. 
In this way we therefore created $9\times3$ synthetic pulse profiles to be analysed, see Figure \ref{fig:simulated} for an example.

\begin{table*}
\begin{tabular}{llll}
    \hline
    \textbf{Parameter} & \textbf{Priors for inference} & \textbf{Input values for synthetic data} & \textbf{\citealt{Psaltis2014}} \\
    \hline
    Mass ($M$) [$M_{\odot}$]& $M\sim U(1,3)$ & $ 1.6$ & $ 1.6$ \\
    \hline
    Radius ($R_{\mathrm{eq}}$) [km] & $R_{\mathrm{eq}}\sim U(3r_g(1),16)^a$& $10$ & $10$ \\
    \hline
    Compactness condition & $R_{\mathrm{eq}}/r_g(M)^a >3$& &  \\
    \hline
    Distance ($D$) [kpc] & $D\sim N(0.3,0.009)^b$ & $0.3$ & N/A \\
    \hline
    Observer inclination ($i$) [$^{\circ}$] & $\cos\left(i\right)\sim U(0,1)$& $30$ & $30$ \\
    \hline
    Hot spot phase shift ($\phi$) [cycles]& $\phi\sim U(0.25,0.75)$& $0$ & N/A \\
    \hline
    Hot spot colatitude ($\theta$) [$^{\circ}$]& $\cos\left(\theta\right)\sim U(-1,1)$ & $40$ & $40$ \\
    \hline
    Hot spot angular radius ($\zeta$) [$^{\circ}$]& $\zeta \sim U(0,90)$& $5$ & $\lesssim 10$ \\
    \hline
    Hot spot effective temperature ($\log_{10}\left({T\left[\mathrm{K}\right]}\right)$) & $\log_{10}\left({T}\right)\sim U(5.1,6.8)$& $6.1$ & Not specified \\
    \hline
    \NICER XTI energy-independent scaling factor ($\alpha_{\mathrm{XTI}}$) &$\alpha_{\mathrm{XTI}}\sim N(1,0.1)^c$ &  $1$ & N/A \\
    \hline
    Column density ($N_\mathrm{H}$) [$10^{20}\,\mathrm{cm}^{-2}$]& $N_\mathrm{H}\sim N(2.3, 1.6)^d$& $2.3$ & N/A \\
    \hline
\end{tabular}
\caption{All model parameters used for simulation of synthetic X-ray pulses in this work, the corresponding prior used for inference, and the corresponding value adopted in \citet{Psaltis2014}. The hot spot temperature is marked for \citet{Psaltis2014} as {\it Not specified} rather than {\it N/A}, because a finite temperature is assumed but not explicitly disclosed. \\
$^a$: where $r_g$ is the gravitational radius $r_g = GMc^{-2}$\\
$^b$: limited to $\pm 10\,\sigma$\\
$^c$: limited to the range [0.7,1.3]\\
$^d$: limited to the range [0.1,5.0].}
\label{tab:Parameters}
\end{table*}

\subsection{Inferences}
\label{subsec:inferences}
To assess the dependency of radius uncertainties on the \ac{MSP} spin frequency, we set up an \XPSI inference run for each of the 27 synthetic pulse profiles previously generated. All of these runs were performed assuming the same \ac{NS} surface pattern model adopted for the generation of the synthetic pulse profiles (i.e.  a single circular hot spot emitting at a uniform temperature). 
These inference analyses also give us the opportunity to check whether model parameters are reasonably recovered. 
We run these parameter estimations on the local cluster HELIOS\footnote{Operating system: CentOS 7; Model name: AMD EPYC 7452 32-Core Processor; Base frequency: 2.35 GHz).}. 

As mentioned above, \XPSI has been most commonly used in combination with \MultiNest. 
With this sampler we explore the model parameter space and compute the desired Bayesian posterior  distributions.
Initialising \MultiNest requires fixing some settings; in particular \citet{Vinciguerra2023a, Vinciguerra2023b} pointed out the impact of sampling efficiency (the inverse of the hypervolume expansion factor), evidence tolerance (i.e. the stopping criterion) and number of live points.  
In this work we always fix them to 0.3, 0.1 and 5000 respectively. 
We do not use the \MultiNest multi-mode or mode separation modality, which was adopted in a few instances by \citet{Riley2019, Vinciguerra2023a, Vinciguerra2023b}.

\section{Results}
\label{sec:results}
Here, we report our findings. 
First, in Section \ref{subsec:Precovery}, we focus on checking whether the inferred posterior distributions are consistent with expectations, given the injected parameter values. 
Then, in Section \ref{subsec:Rvsf}, we present how the inferred relative uncertainties on the \ac{MSP} radius vary with the pulsar spin frequency, and compare this trend with the relations presented in Section \ref{subsec:prediction}. 

\begin{figure}
	\includegraphics[width=9cm]{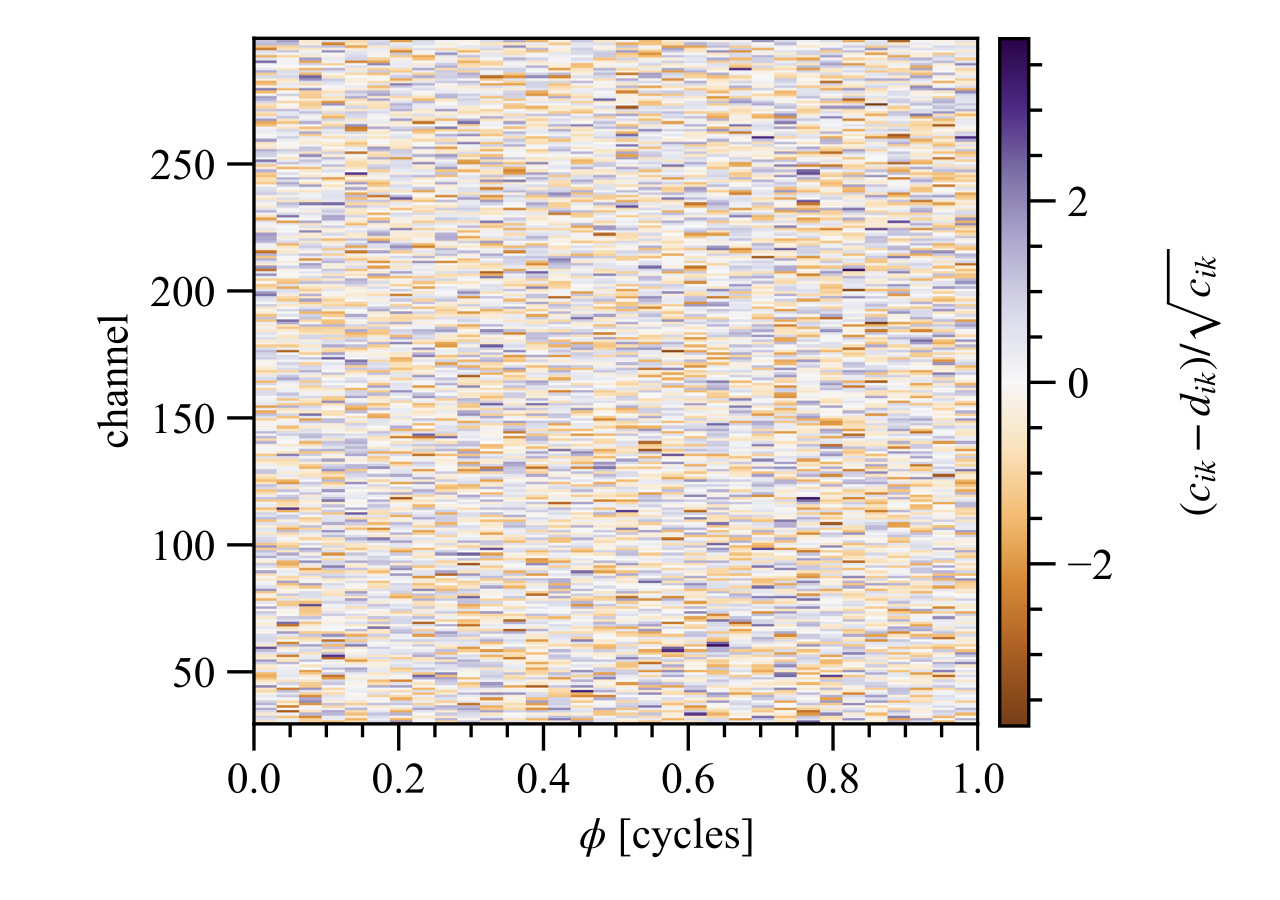}
    \caption{Example of a residual plot for data simulating the emission of a pulsar rotating at 600\,Hz. The considered PI channels are displayed on the y-axis, while in the x-axis we report the adopted bin in rotational phase. The colour bar shows the residuals corresponding to each channel-phase bin. The residuals are calculated according to the formula on the label, where $c_{ik}$ and $d_{ik}$ stand respectively for counts from the inferred signal and from the synthetic data. The inferred signal is an average over 200 samples randomly picked from the posterior distributions (see Figure 13 of \citealt{Riley2019} for further details). 
    }
    \label{fig:residual}
\end{figure}

\begin{figure*}
	\includegraphics[width=2\columnwidth]{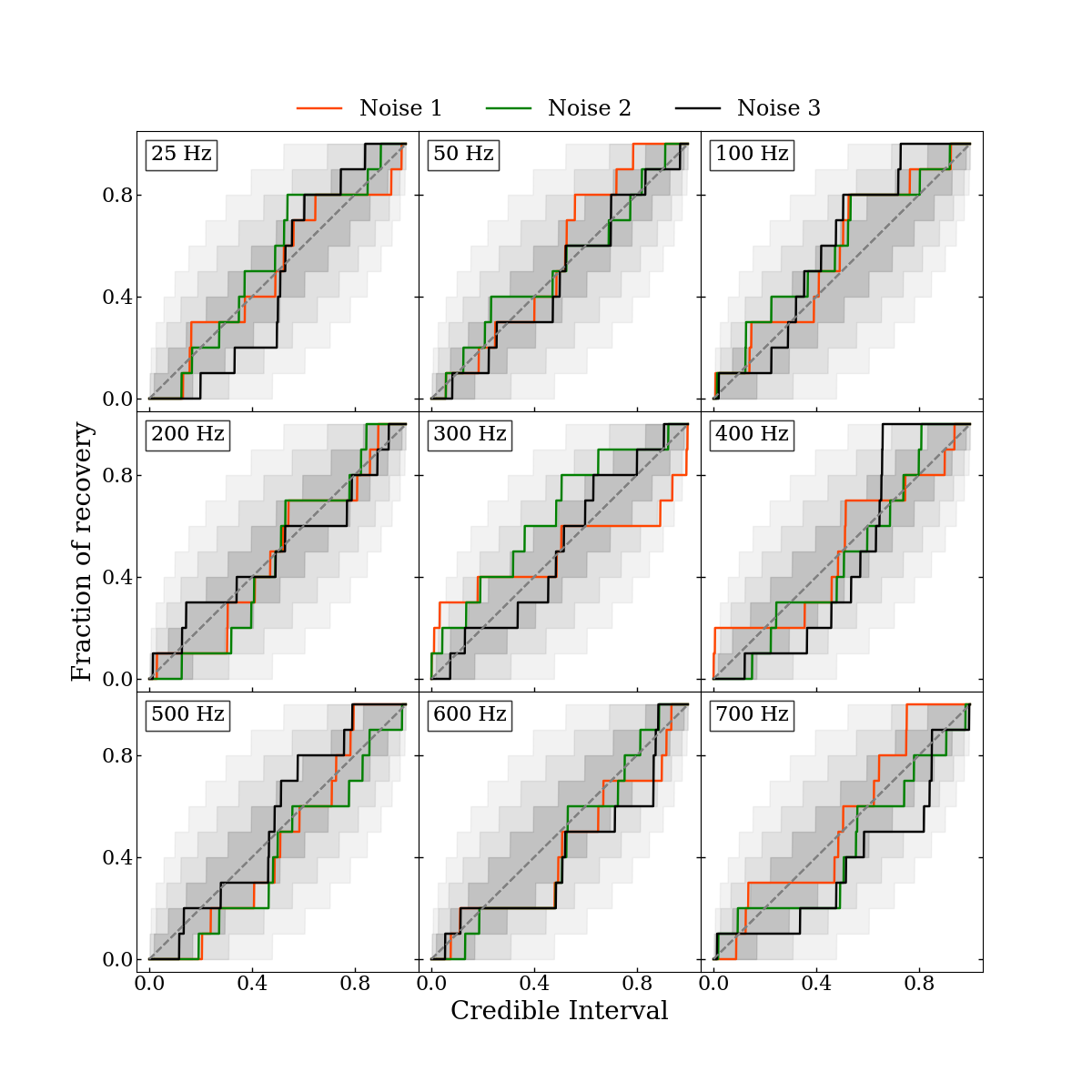}
    \caption{Probability-Probability (pp) plots for each pulsar spin frequency: from the slowest on the top left to the fastest on the bottom right corner. 
    Each of these panels shows how often (on the y-axis) the injected value is recovered at a certain percentile (on the x-axis) of the inferred posterior distribution. To boost the available statistics, all of the model parameters are included in each of the represented curves. Note that since in this way we do not account for degeneracies and correlations present between the different model parameters, this Figure only provides a superficial first insight (more specifically if the curves in this Figure had been far from expectations, this would have flagged the presence of significant biases; the fact that our findings are consistent with expectations, however, does not prove the absence of biases). 
    Each pp plot contains three lines, representing the results for each of the three inference runs we performed per spin frequency, analysing data produced with different noise realisations (note that all of the random seeds adopted to generate the different pulse profiles are different from each other). 
     }
    \label{fig:PP_plots}
\end{figure*}

\begin{figure*}
	\includegraphics[width=1\columnwidth]{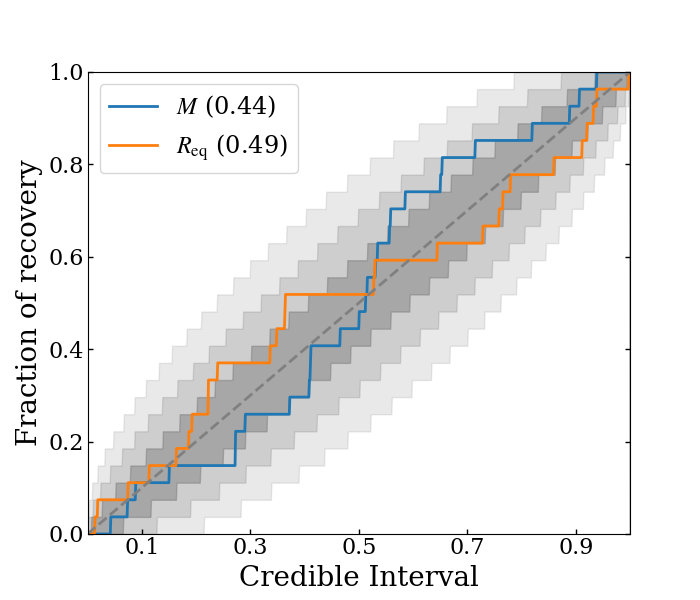}
    \caption{Probability-probability plot of the mass and radius, combining the results obtained from all our inferences. They include analyses for data produced with different pulsar spin frequencies and noise realisations. The values in brackets are the p-values, evaluating the consistency between the recovered curves and theoretical expectations. Since for both mass and radius this value is well above the 0.05 threshold, the null hypothesis (of the data points being drawn from a uniform distribution, as theoretically expected) cannot be rejected. }
    \label{fig:mass_radius_PP_plot}
\end{figure*}

\subsection{Parameter Recovery}
\label{subsec:Precovery}
To examine whether the solutions found in the inference process fit the data well, we use \XPSI to generate residual plots. 
These show the difference between the data and the model prediction, conditioned on 200 samples randomly selected from the inferred posterior. 
To understand the deviation compared to expectations, we weight the difference in each phase-channel bin by the predicted standard deviation $\sigma$. 
If the only difference between the data and the pulse profile simulated in our inference analysis is the presence of Poisson noise, the residual plot will show no significant pattern and deviations per bin will be limited to a few $\sigma$. 
This is what we observe for all of our $3\times 9$ simulated data sets, i.e. independently of the assumed spin frequency: with its inference process, \XPSI finds parameter vectors that adequately represent the data, see e.g. Figure \ref{fig:residual}.

Since data can be differently sensitive to different model parameters, this does not necessarily correlate with the performance of our inference. 
For a rough evaluation of the goodness of the posterior distributions recovered by our analyses, we look at the probability-probability (pp) plots \citep[see e.g.][]{Berry2015,Krishna2023,Kini2023a,Kini2023b}. 
These count (and later normalise), for all cases considered, how many injected values are recovered at different percentiles of the posterior distributions. 
Since here we are interested in how our parameter estimation process is affected by the \ac{NS} spin, we build pp plots for each frequency. 
Ideally one should look separately at each parameter. However for such a plot to be significant (i.e. for the 1\,$\sigma$ uncertainties not to spread over most of the plot), one would need considerably more simulations, whereas we only have three.
Instead we have grouped together the different model parameters to check whether, as a group, their corresponding posterior distributions behave according to statistical expectations. 

This is represented in Figure \ref{fig:PP_plots}; where each noise realisation for a given frequency is shown with a different curve. 
All of our inference runs perform, in general, as expected: most of their pp plots are lying within $\pm1$ or $\pm2$ $\sigma$ (two darker gray areas)\footnote{These bands represent the uncertainties arising from the limited and, for statistical purposes, relatively small number of model parameters. The $\pm1$ ($\pm2$) $\sigma$ range is determined by the 16th and 84th percentiles of the percent point function of a binomial distribution, which characterizes a sample size of 10 (the number of inferred parameters per run for the emitting model) and a success rate of 68\% (95\%) (the credible interval of interest).
} from the ideal diagonal case. 
It should be noted, however that the parameters of our inference are not completely independent from each other and hence these plots should be taken as indicative. 
In our analyses, we note the absence (when considering all 27 runs) of parameters which always have the injected value on one specific side of the posterior distribution. 
However there are some clear correlations: the lower the inferred radius, the higher the estimated hot spot temperature and the larger its size; while the strong sensitivity of our analyses to the compactness parameter dictates that inferred higher masses are usually associated with large radius values. Because of their strong effects on the photon energy, both the radius correlations with mass and temperature are quite smeared out for low \ac{NS} spin frequencies (i.e. for frequencies of 50 and 25\,Hz). 
Moreover, for distance and \NICER energy-independent scaling factor, it is clear that our analyses currently overestimate the statistical uncertainties, as their injected values always lie 
extremely close the median value compared to the whole posterior distribution.  
Finally we also notice that, in all our posteriors, there is a bimodality in the hot spot colatitude, which significantly impacts the meaning of the corresponding 68\% credible interval.

Since the recovery performance does not seem to be affected by the spin frequency (see pp plots of Figure \ref{fig:PP_plots}), we grouped all of our 27 inferences to check the recovery of single model parameters;
since the main target of \ac{PPM} is constraining the equation of state, we focus on masses and equatorial radii. 
It should be noted that our tests focus on a single point within the parameter space, meaning we consider only one set of true parameter values to construct our synthetic data. 
Within this limitation, Figure \ref{fig:mass_radius_PP_plot} shows, once again, that our analysis is effective and unbiased in the estimate of \ac{NS} masses and radii, since both of the curves lie well within the $2\sigma$ area, and when compared with the theoretical case we obtain quite high p-values (see legend of the same Figure). 

\subsection{Radius uncertainty dependence on pulsar spin frequency}
\label{subsec:Rvsf}
\begin{figure*}
	\includegraphics[width=2\columnwidth] {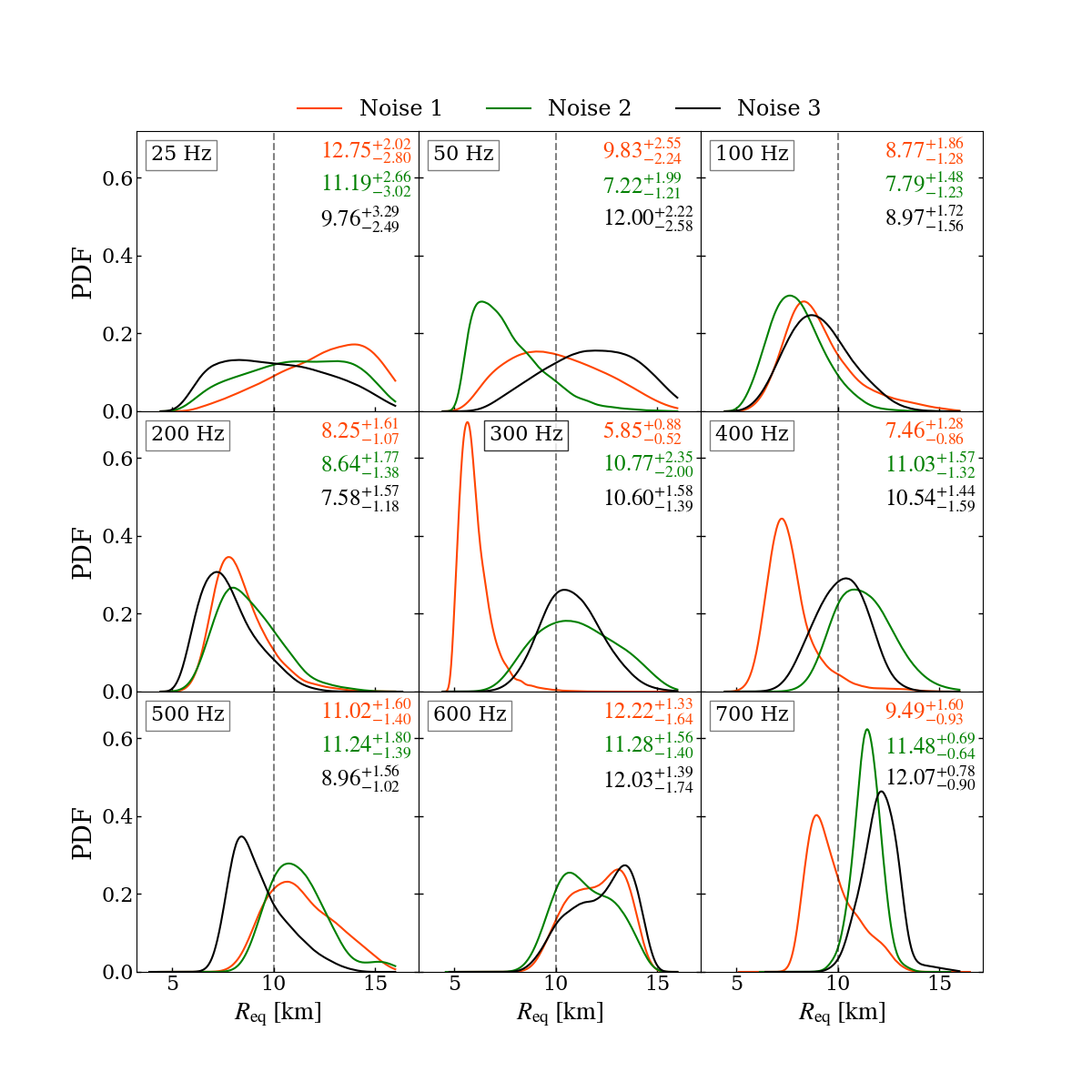}
    \caption{Inferred posterior distributions for the equatorial radius $R_\mathrm{eq}$. Similarly to Figure \ref{fig:PP_plots}, they are grouped for different pulsar spin frequencies going from the slowest on the top-left panel to the fastest on the bottom-right (as also shown in the plot legend). Each plot contains the posteriors inferred from the analyses of the three data sets constructed on different Poisson noise realisations and the injected value, marked with the vertical dashed line. The numbers reported on the top-left corner of each panel represent the 68\% credible interval from the median value. Note again that all the noise realisations are built on different random seeds. }
    \label{fig:radiusposterior}
\end{figure*}
To test the impact of spin frequency on our inference analyses and to compare to the predictions outlined in Section \ref{subsec:prediction},  
 we now focus on the posteriors of the \ac{NS} radius estimated in our case study.  We report the $R_{\mathrm{eq}}$ distribution inferred for each of our simulated data sets in Figure \ref{fig:radiusposterior}. 
If the \ac{NS} rotates with a spin frequency $<100\,$Hz our constraining power is clearly poor: as shown in the upper-left and upper-center plots of Figure \ref{fig:radiusposterior}, our posterior distributions span with meaningful probabilities all or most of the prior range\footnote{In the only instance (noise 2, $50\,$Hz) where this is less so, the peak clusters around a value considerably different from the truth. }. 
The posterior plots corresponding to higher \ac{NS} spin frequencies show considerably more peaked distributions.
This behaviour is qualitatively in line with expectations (including from previous studies, \citealt{Psaltis2014,MillerLamb2015,Bogdanov2021}). 

\begin{figure*}
	\includegraphics[width=2\columnwidth]{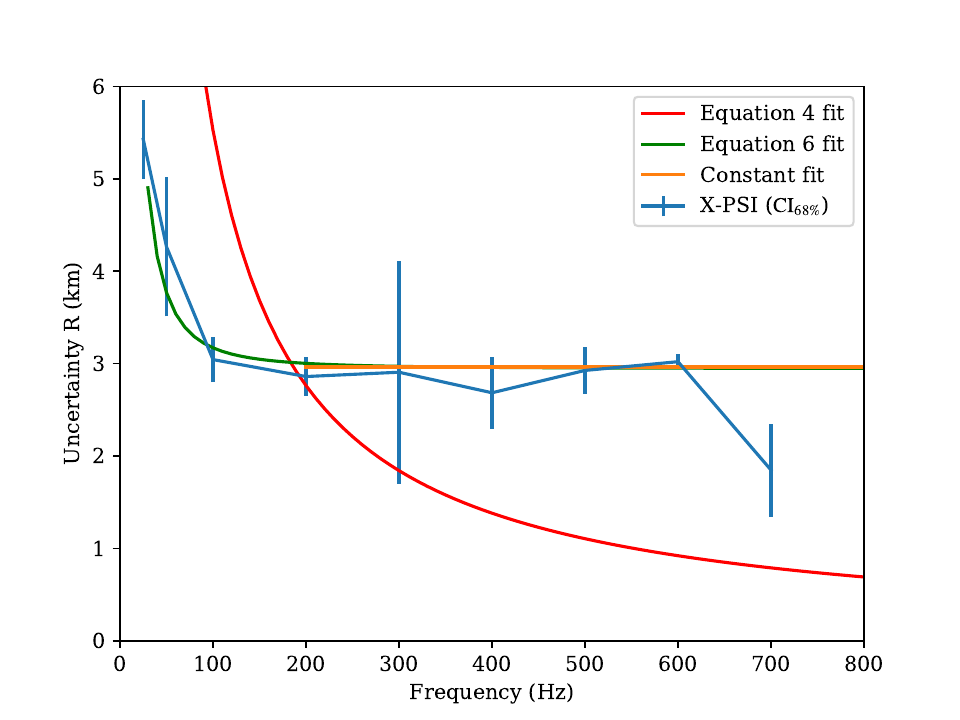}
    \caption{Representation of the radius uncertainty as a function of the pulsar spin frequency.  In blue we plot the uncertainties obtained with our inferences. The vertical bars represent the $\pm1$ standard deviation over the three inferred 68\% credible intervals over the $R_\mathrm{eq}$ posteriors. 
    In red and green we show the fit to the data points, according to Equations \ref{eq:PsaltisFit} and \ref{eq:deltaR_alpha_beta}, respectively.  Note: we have also checked the exact values of the predicted uncertainties from Equation 4 of \citet{Psaltis2014} (our Equation \ref{eq:relR}) and find that they are consistent with the values inferred from our fit using Equation \ref{eq:PsaltisFit}.} 
    \label{fig:scaling_relation}
\end{figure*}

Figure \ref{fig:scaling_relation} shows quantitatively how the radius uncertainties depend on the \ac{NS} spin frequency. 
Here we compare our findings (in blue) with the predictions from Equations \ref{eq:PsaltisFit} and \ref{eq:deltaR_alpha_beta} (see Section \ref{subsec:prediction} for more details). 
 This Figure shows that in this restricted case study, the radius uncertainties are not simply inversely proportional to the \ac{NS} spin frequency, as in Equation \ref{eq:PsaltisFit} (the red curve poorly fits the blue data points, the best fitting $\alpha$ value was $\sim 552\,$km\,Hz). 

The general form of the function expressed by Equation \ref{eq:deltaR_alpha_beta} seems however to be able to explain our empirical findings, suggesting, as hinted at by Equation \ref{eq:DRR}, that other parameter uncertainties also contribute to the scatter of the inferred radius posterior. 
The values of the $\beta$ and $\gamma$ parameters obtained from this fit are: 8.67\,km$^2$ and $1.38\times 10^4$\,Hz$^2$\,km$^2$. 
We note that this form, at least for the specific parameter vector injected to create our data sets, flattens out quite quickly, showing no significant gain when considering particularly fast \acp{MSP}; spin frequencies of 
$\sim 200\,$Hz seem to already reach the plateau of our empirical function. 
Computing $\chi^2$/dof for data points $\ge 200\,$Hz reveals that indeed a constant is favoured compared to Equation \ref{eq:deltaR_alpha_beta} (even when this is fitted to the same reduced set of data). 

\begin{figure}
\centering
	\includegraphics[width=0.7\columnwidth] {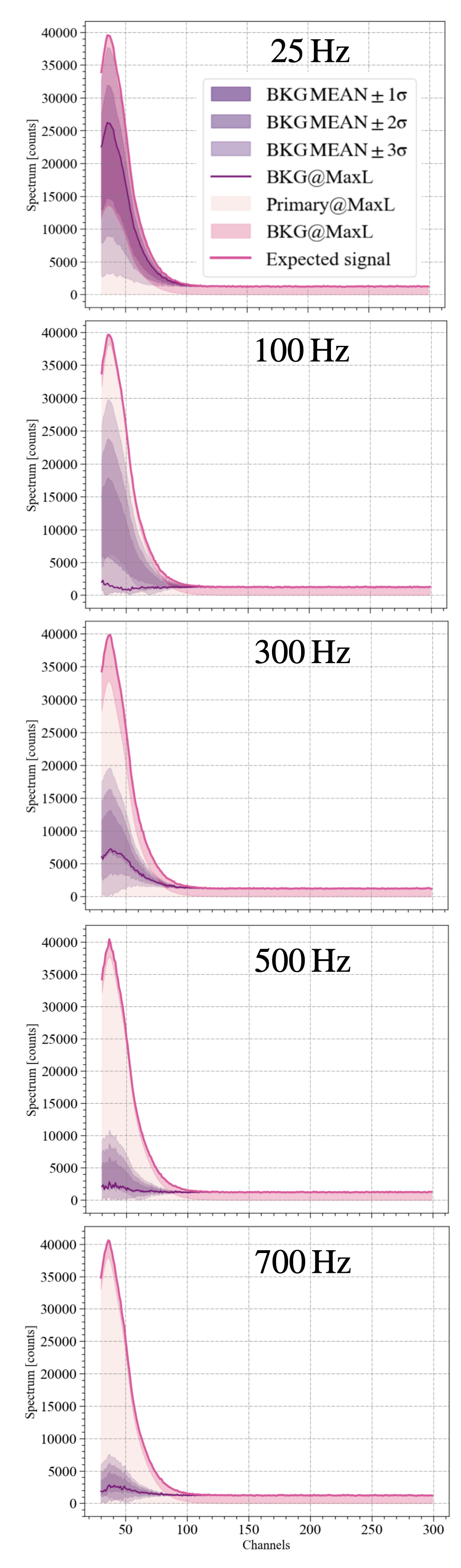}
   \caption{Schematic representation of how inferred background changes with the \ac{NS} spin frequency, for a sub-sample of our simulations. The legend in the top panel applies to all:  light pink areas show the contribution of the primary hot spot;  above this lies background (darker pink) for the maximum likelihood sample. The pink solid line shows the expected signal, and the purple line the background associated with the maximum likelihood sample. The purple areas show the mean $\pm 1\sigma$, $\pm 2\sigma$ and $\pm 3\sigma$ going from darker to lighter shades. The injected background value corresponds roughly (except for noise) to the flat count rate visible at the highest PI channels. }
    \label{fig:BKGchange}
\end{figure}

Interestingly we notice that the uncertainties over the inferred background, as well as its estimated value\footnote{As in the \NICER MSP \ac{PPM} analyses, we consider as background all the counts that are not directly attributable to the thermal emission of the hot spots.} seem to be strongly related to the \ac{NS} spin frequency; the shorter the rotation period, the better we constrain the background. 
This is shown clearly in Figure \ref{fig:BKGchange}. In \citet{Psaltis2014}, the background is assumed to be known, which may explain some of the differences. 

Looking at Figure \ref{fig:radiusposterior}, we notice that for spin frequencies of $100\,$Hz and $200\,$Hz, our inferred posteriors systematically underestimate the true value of the \ac{NS} radius\footnote{This was not apparent in Section \ref{subsec:Precovery},  since then we were looking at all spin rates.} and show very small variations, despite the random processes acting in our inference procedures, and the different noise realisations adopted to create the analysed data sets. 
The injected $R_{\mathrm{eq}}$ value is instead well recovered by two of the three inferences for the \ac{NS} spin frequencies of $300\,$Hz, $400\,$Hz and $500\,$Hz. For faster stars we notice a slight tendency of overestimating the true value. 
These opposing trends at high and low spin rates may compensate each other in the pp plot representation of the radius (orange curve in Figure \ref{fig:mass_radius_PP_plot}). A simple $\chi^2$ test on the median values of the posteriors confirms that the limited data available hint at a positive correlation between the inferred radii and the spin frequency of the \acp{MSP}. 
We defer further investigations of this trend to future studies. 

We notice that the resemblance between the three posteriors for each individual spin frequency is variable and does not seem to scatter randomly around the truth. This diverse behaviour of the posterior distributions is due to the Poisson noise realisation. There are many possible model parameter combinations that can generate pulse profiles similar to those expected from our injected values, which in turn can be attributed to the 
many correlations and degeneracies between the model parameters (for example emission constant in rotational phase could be generated from the background, the compactness or even geometrical properties such as a hot spot present on the pole or very low inclination angles).
The particular noise realisations then determine which of these parameter combinations better match the considered data set and hence the associated likelihood surface. 

The likelihood surfaces are of course also affected by the details of the analysed pulse profile. 
Since the expected pulse profile exhibits different features depending on the \ac{NS} spin frequency, different groups of solutions are then more or less supported by the data, once Poisson noise has been added (see also \citealt{Vinciguerra2023a} for more detail on the impact of noise realisations). 
This is also probably why the behaviour of the three posteriors is more similar for similar spin frequencies. 
Overall, higher frequencies seem to push the posteriors to slightly higher radius, with the exception of a distinct bulk of solutions, associated with considerably smaller radii. 
This set of solutions, for which the median value also increases with the \ac{NS} rotational speed, begins to be present at 300\,Hz and disappears at 600\,Hz.  
Of course, given only three noise realisations per spin frequency, we cannot exclude that similar solutions could have also been found at lower and higher rotational speeds with a larger number of noise realisations; it indeed seems to reappear at spin frequency of 700\,Hz.  In addition, although the posterior distributions sometimes peak quite far away from the injected value, the likelihood $\mathcal{L}$ associated with the injected parameter vector is always lower than the maximum likelihood found in the sampling process, sometimes considerably so (for $f=300\,$Hz the difference reaches a factor of $\sim8$ in $\log_e\mathcal{L}$).

\section{Discussion}
\label{sec:discussion}

\subsection{Reliability checks}
Firstly, we performed tests to confirm the reliability of our analyses. By inspecting the residuals of each of our inference runs, as done for \NICER \ac{PPM} inferences, we made sure that the solutions found represent our simulated data well. As an additional test, we produced pp plots for each of the 9 \ac{NS} spin frequencies tested, considering all of the model parameters: we do not observe any particular trend, and the curves seem to follow statistical predictions. 
Moreover the mass and radius pp plot, considering the results of all 27 inferences, is in complete agreement with expectations. 
Nevertheless, a detailed look at the radius posteriors (see Figure \ref{fig:radiusposterior}) shows 
a slight tendency for increasing values of the inferred radius for higher spin frequencies.

\subsection{The radius uncertainties do not scale with $\mathbf{1/f}$}
    The main results are shown in Figure \ref{fig:scaling_relation}:
a simple inverse proportionality between radius uncertainty and spin frequency, as expressed by Equation \ref{eq:PsaltisFit}, cannot fit our data points. However a more comprehensive development of Equation \ref{eq:amplitude_ratio} - as in Equation \ref{eq:deltaR_alpha_beta} - that can take into account the other inferred parameters and their respective uncertainties, arriving at a slightly different frequency dependence, fits our data points quite well. 

\subsection{Flattening above $\mathbf{\sim200\,}$Hz}
Our fit predicts very wide uncertainties for frequencies below $\sim100\,$Hz (as is clear from Figure \ref{fig:PP_plots}), 
and an almost flat curve for spin frequencies higher than $\sim 200$\,Hz. 
Follow-up studies would be needed to determine whether this inflection point of the fit is a general feature of \ac{PPM} analyses, or whether it depends on the specific emitting surface pattern chosen to simulate the data. 
If this finding is general, it would have important implications for future missions and their observing plans: 
most \acp{MSP} could be considered interesting candidates for \ac{PPM} techniques and priority should be given to targets based on properties other than their spin frequency.

\subsection{Comparison with \citet{MillerLamb2015}}
\citet{MillerLamb2015} explored how mass-radius inferences would be affected by approximations in the waveform and a few changes in the model parameters. 
In particular they consider two identical cases (with a much more favourable configuration compared to our case, with $\theta = i = 90^{\circ}$), where only the \ac{NS} spin frequency was different: in one instance the \ac{NS} spins at a rate of 300\,Hz and in the other at 600\,Hz. 
In this study they reached overall similar conclusions to us, as they discourage the direct use of spin frequency for prioritising targets and instead suggest pre-analyses. 
However, unlike in our analysis, they reported considerable improvements when increasing the spin frequency from $300$\,Hz to $600$\,Hz. 
This difference could be due to several factors:
\begin{itemize}
    \item the different parameter vectors adopted in their simulations  (higher radii and more favourable inclinations and hot spot colatitudes) which increase the hot spot velocity along the line of sight and enhance the pulsating component of the emission;
    \item the eclipsing of the hot spot, present (as a consequence of the different choices of injected parameter vectors) in their simulations but not in ours;
    \item the number of model parameters: 10 in our case, 5 in theirs;
    \item the assumption in their simulations that either the distance or the redshifted temperature of the hot spot is known; 
    \item the background: they use a higher level of background and describe it with 31 nuisance parameters compared to our 269;
    \item the treatment of background parameters: in our case we marginalise over them, while they maximise the likelihood over them; 
    \item the inclusion in our analysis of the \NICER response matrix; 
    \item other differences in the pipeline, including e.g. the adopted sampler.
\end{itemize}

\subsection{Comparison with \citet{Bogdanov2021}}
\citet{Bogdanov2021} presented several simulations and inferences aimed at cross checking the mass-radius results obtained from the analysis of \NICER data, using both an earlier version of \XPSI and the Illinois-Maryland ray-tracing code. 
In one instance they considered the emission from a 600\,Hz \ac{MSP}, of mass $M = 1.4\,M_\odot$ and radius $R_\mathrm{eq} = 10\,$km characterised by a single hot spot. For this case they found much smaller radius uncertainties and more accurate estimates than we find in our study (the \XPSI run in that paper, for example, yielded a radius of $12.12^{+0.43}_{-0.41}\,$km).  There are several reasons that likely contribute to this difference; in particular the hot spot in the \citet{Bogdanov2021} case is located at the equator, with a viewing angle of 90$^\circ$, offering a much more favourable configuration.  Moreover the \ac{NS} is placed at a distance of 0.2\,kpc (compared to our 0.3\,kpc) and in \citet{Bogdanov2021} the simulation does not include any interstellar absorption. Other differences include the version of the \NICER response matrix used, the instrument channel range considered, and the fact that only a single noise realisation was tested.

\citet{Bogdanov2021} also presented evidence for improvements in credible intervals as spin frequency increased, based on a comparison carried out using the Illinois-Maryland ray-tracing code.  They examined the impact on the mass-radius contours when considering the emission of \acp{NS} rotating at 1\,Hz and at 600\,Hz, holding the injected values of the other parameters constant. They considered two different hot spot sizes and in both cases, the fastest rotating \ac{NS} gave rise to much better constraints. This is qualitatively in agreement with our results, since we also observe large improvements across this large range of rotation rates.

\subsection{Interpretation: the impact of an unknown background}
When comparing our results with those presented in \citet{Psaltis2014}, it is important to note that we assume no meaningful background constraints. 
In \citet{Psaltis2014}, on the other hand, the background is assumed to be known. As outlined in Section \ref{sec:results}, this different approach to the background may have a significant impact on our results. 
Figure \ref{fig:BKGchange} shows that the most prominent effect of higher \ac{NS} spin rates, in our data analysis framework, consists of a better estimate of the background counts. However our study also reveals that this does not directly translate to narrower inferred radius posteriors. 

While for future missions our background treatment (i.e. assuming no constraints) is quite pessimistic, the background level and shape (constant) that we simulated represent what is for current analyses a best case scenario. For all \NICER sources for which \ac{PPM} analysis has been published, the background is neither small (compared to the hot spot component) nor the same in all PI channels. 
Future studies should therefore check whether a more realistic choice would affect our findings \citep[see also][on the importance of background constraints]{Ozel16,Salmi2022};  higher background counts would imply larger statistical fluctuations that could increase the complexity of the parameter estimation analysis. However we do expect situations in the future where the background will be very tightly constrained and even lower than we assumed; for example for NewAthena \citep[][]{NewAthena}.

\subsection{The complexity of realistic analyses}
In our inference analyses we estimate the posterior on a 10-dimensional parameter space. 
This is a fairly complex process which is further complicated by degeneracies and correlations between the model parameters. 
It is not uncommon to encounter a multi-modal structure in the posterior distribution. 
This is, for example, the case for all of our estimations of $\theta$. 
As expected \citep[][]{ViironenPoutanen2004}, we also observe a strong correlation between the colatitude of the hot spot and the observer inclination.  Degeneracies and multi-modalities would need to be considered if we wanted to calculate the expected radius uncertainties from Equation \ref{eq:DRR} (see last paragraph of Section \ref{subsec:caveats}).

\subsection{Caveats}
\label{subsec:caveats}
\begin{description}
    \item[{\bf Specificity of the analysed case and expectations.  }]
The main limitation of our findings concerns the specificity of the case considered in this preliminary study. It remains to be determined whether Equation \ref{eq:deltaR_alpha_beta} could well fit the inferred radius uncertainties for other pulse profiles, built on other parameter vectors, and whether the inflection point of the fitting function would remain at similar, relatively low, spin frequencies. 
For example, an increased sensitivity to the spin frequency is expected whenever the hot spot velocity projected on the line of sight increases and this could push the flattening to larger spin frequencies.  
Moreover, if we assumed that our fitting formula (Equation \ref{eq:deltaR_alpha_beta}) is a representation of Equation \ref{eq:DRR}, it would be reasonable to expect that tighter priors on the model parameters $\theta$ and $i$\footnote{In the ideal case where the uncertainties are there zero, we indeed go back to Equation \ref{eq:PsaltisFit} and we should therefore expect no significant flattening.} would reduce the value of the fitted $\beta$ and therefore let the frequency dependence emerge more prominently, maybe moving the flattening point. There are also other more complex cases that should be considered, e.g. the addition of a second spot and more complex emission patterns.
In addition, in a more realistic atmosphere model the beaming and thus the pulse shape would depend on the surface gravity (which varies as $M/R_{\mathrm{eq}}^{2}$). 
This could improve the accuracy and precision of the radius measurement, since information about the radius would then be derived from more than just the ratio of the harmonic to the fundamental.

    \item[{\bf Possible frequency dependence of  uncertainties.}] 
    Equation \ref{eq:deltaR_alpha_beta} is able to fit our data points well. 
    This fitting formula was inspired by Equation \ref{eq:amplitude_ratio} and developed into Equation \ref{eq:DRR}, under the assumption that only the relative uncertainty of $C_\mathrm{2}$
  depends on frequency. However this may not be a well-founded assumption; more complex frequency dependencies may turn out to be needed if one were to explore a wider parameter space. Nevertheless for our restricted case, according to a simple $\chi^2$ test, Equation \ref{eq:deltaR_alpha_beta} contains sufficient complexity in the spin frequency range (25-700)\,Hz. 
    We have checked the posterior distributions of $\theta$ and $i$, finding no clear frequency dependence of their width. 
     We also note that our fitting formula would allow relative uncertainties on other parameters to depend on the spin frequency if they were inversely proportional to it. 
    \item[{\bf The effect of different noise realisations.}] Our data shows considerable scatter, highlighting that theoretical predictions based solely on source properties and exposure time may not be sufficient to make accurate estimates of uncertainties, in contrast to what was previously thought \citep{Watts2016,Watts2019}.
    \item[{\bf Statistical caveats.}]  
The low number of inference runs per spin frequency is a limitation on the robustness of our findings. As shown by the blue vertical bars in Figure \ref{fig:scaling_relation}, the scatter in the posterior width can be considerable and the ones that we report here may not be representative. We do however find that the width of the radius posterior does not vary significantly with \ac{NS} spin for most of our simulations above $200\,$Hz (with the exception of only the 700\,Hz case) i.e. over a larger number of trials.
    \item[{\bf Treatment of radius uncertainties.}] 
    To derive the formulas presented in Section \ref{subsec:prediction} we expanded the derivative of Equation \ref{eq:amplitude_ratio}, assuming that uncertainties are measured in standard deviations.
    In general however in \ac{PPM} uncertainties are expressed in credible intervals (most commonly the 68\%, or alternatively the 90\%, 95\% or  99\%). One standard deviation would be half the 68\% credible interval \footnote{Note that constant factors in front of the standard deviation cancel each other out as long as they are used for all the parameters accounted for in the considered formula.} when the distribution takes the shape of a Gaussian; as shown in Figure \ref{fig:radiusposterior}, the posterior often approximates a Gaussian, however there are some instances where it is significantly different. 
    \item[{\bf Hypothetical application of Equation \ref{eq:DRR}.}]
    If we were to exactly compute Equation \ref{eq:DRR} with inferred quantities, we would have derived unrealistically large radius uncertainties. 
    This is likely due to the complexity of the real problem at hand, which includes exploring a 10 dimensional parameter space in contrast with the 4 dimensions considered in \citet{Psaltis2014}. 
    Despite the huge difference in normalisation\footnote{
    The discrepancies between the $\beta$ and $\gamma$ parameters that we obtain from the fits to the ones calculated according to Equation \ref{eq:DRR} can reach factors of $\sim 50$, likely due to parameter correlations.}, we observe a frequency dependence of the right hand side of Equation \ref{eq:DRR} similar to the one we derive for the inferred $\Delta R_{\mathrm{eq}}$: a considerable drop up to $\sim 100\,$Hz and a flattening from $\sim 200\,$Hz. Analysing the various components entering Equation \ref{eq:DRR}, we notice that the contribution of the relative uncertainty of $C_\mathrm{2}$ dominates up to $\sim 100\,$Hz while for faster \acp{NS} it drops below the level set by $\Delta \theta/\theta$ and $\Delta i/i$. If $\Delta C_{\mathrm{2}}/C_{\mathrm{2}}$ is indeed the only relevant term depending on the spin frequency that contributes to the inferred radius credible intervals, this would also explain the observed inflection point. 
    However there are caveats and complications involving the exact estimate of Equation \ref{eq:DRR}: we are considering 68\% credible intervals instead of standard deviations; the posteriors of $\theta$ and $i$ are highly correlated, which would modify Equation \ref{eq:DRR} and reduce the combined uncertainty; the inferred 1D posterior of $\theta$ always shows a bimodality, this also artificially increases $\Delta \theta$; Equation \ref{eq:amplitude_ratio} assumes a specific spectrum of the emission (according to Equation 66 of \citealt{Poutanen2006}) which does not clearly fit the blackbody case considered in our study; and finally, if we were to infer $C_\mathrm{1}$ and $C_\mathrm{2}$ from Fourier analyses of our synthetic data sets, we would include the background in  our calculations and hence derive different values compared to what is assumed in \citet{Psaltis2014}\footnote{We do indeed find mismatches that can reach 100\% between the harmonic amplitudes estimated through Equation \ref{eq:amplitude_ratio} and Fourier analyses of the simulated signals. }.  
\end{description}

\section{Summary}
\label{sec:conclusions}
In this paper we investigated the dependence of the equatorial radius uncertainties (the posterior credible intervals obtained from \ac{PPM} analyses), on pulsar spin frequency.  
This is crucial to prioritize \ac{PPM} targets in accordance with their \ac{EoS} constraining power. 
Our findings should therefore inform the observing strategies applied to current and future missions that aim to carry out \ac{PPM} analyses such as \NICER, STROBE-X, eXTP and NewAthena. 

We test this dependency in a realistic framework, adopting for the inference analyses the same procedure and software (\XPSI) currently being used to perform \ac{PPM} analysis of \NICER data.  We focus on the theoretical case developed by \citet{Psaltis2014}, since these make analytic predictions that can be tested. 
With \XPSI we simulate the expected pulse profile generated by X-ray pulsars, spinning at 9 different frequencies spanning the range (25-700)\,Hz and emitting from a single circular hot spot with uniform temperature (with parameter values as described in Figure \ref{fig:parameters} and reported in Table \ref{tab:Parameters}). 
To each of these 9 pulse profiles, we apply 3 random realisations of Poisson noise, to obtain 27 simulated profiles, as they might currently be observed by \NICER.  We then analyse each of these synthetic observations to infer a posterior distribution for each of the model parameters. 
We have confirmed the adequacy of our analysis via residual and pp plots. 

We then study the dependency of the inferred radius credible interval on the spin frequency of the simulated pulsars.  A simple inverse proportionality is not a good fit (see Figure \ref{fig:scaling_relation}). 
We notice large improvements going from 25\,Hz/50\,Hz to the \ac{MSP} regime but then a flattening for \acp{NS} rotating faster than (for this specific case) $\sim 200\,$Hz, likely due to the contribution of other parameter uncertainties. While analytic relations like those developed by \citet{Psaltis2014} are extremely instructive, our study shows that fully understanding the way in which the various parameters (and their priors) contribute to the final uncertainty on the radius requires the kind of Bayesian approach followed here.

We also uncover an interesting difference relating to the background: while our radius posterior is insensitive to the spin frequency of the \ac{NS}, as long as we are  above $\sim$ 200 Hz, the uncertainty in the inferred background improves considerably (see Figure \ref{fig:BKGchange}). 

Taking into account the uncertainties coming from the hot spot colatitude, the inclination and the amplitudes of the fundamental and second harmonics 
gave us some insight into the origin of the observed flattening in the dependency of the radius uncertainty on the spin frequency. The flattening roughly starts when the uncertainties on the hot spot colatitude and inclination dominate over those of the second harmonic. 

The main limitation of our study is that we focus on a very specific set of cases and a limited number of noise realisations: except for the changing spin frequency, we consider a single parameter vector to generate all of our synthetic data sets, and for each case only three noise realisations. 
There may be other hot spot configurations which could display a stronger sensitivity to the spin frequency (e.g. if the projection on the line of sight of the hot spot velocity were larger). For some sources tighter priors may be available, which could also affect the posterior sensitivity to the \ac{NS}  rotation period.
Additional studies are therefore necessary to test how general our findings are. 
If these investigations were to reveal flattening at similarly low spin frequencies, there would be important implications for future missions including \ac{PPM} in their scientific scopes. If general, our results suggest that other source/signal properties may be more useful than pulsar spin to determine target and priorities in current and future \ac{PPM} observing plans. 

\section*{Acknowledgements}
EB carried out this work as part of his Bachelor project for the Bachelor Natuur- en Sterrenkunde and the Masters in Physics \& Astronomy at the University of Amsterdam/Vrije Universiteit Amsterdam.  SV, ALW, YK and TS acknowledge support from ERC Consolidator Grant No.~865768 AEONS (PI Watts). This work was carried out on the HELIOS cluster, exclusively on dedicated nodes funded via this grant.
The authors also thank Martin Heemskerk, Evert Rol and Devarshi Choudhury for helping setting up the analyses on HELIOS and Sebastien Guillot for useful discussions about NewAthena background. 
Finally we thank Dimitrios Psaltis and the anonymous referee for useful comments.

\section*{Data Availability}

All the data, posterior samples as well as the scripts used for the runs and plots are available on Zenodo \citep{zenodoScaling}.

\bibliographystyle{mnras}
\bibliography{main}

\bsp	
\label{lastpage}
\end{document}